\documentclass[12pt,authoryear]{elsarticle}

\usepackage[utf8x]{inputenc} 
\usepackage{mdframed}
\usepackage{eurosym}
\usepackage{amsmath}
\usepackage{amssymb}
\usepackage{bm}
\usepackage{graphicx}
\usepackage{booktabs}
\usepackage{hyperref}
\usepackage{soul}
\usepackage{multirow}
\usepackage[table]{xcolor}
\usepackage{floatpag}
\usepackage{float}
\usepackage{pdflscape}
\usepackage{mathptmx}
\usepackage{soul,color}
\usepackage[round,authoryear]{natbib}
\usepackage{subfig}
\usepackage{mathtools}
\usepackage{setspace}
\usepackage{xcolor}
\usepackage[top=1in, bottom=1.5in, left=1.3in, right=1.3in]{geometry}

\hypersetup{ colorlinks = true, linkcolor = blue, anchorcolor = blue, citecolor = blue, filecolor = blue, urlcolor = blue}
\definecolor{verdone}{rgb}{0,100,0} 
\journal{Journal of Urban Economics - Insights}

{\left\lbrace\begin{array}{@{}l@{}}}%
{\end{array}\right.}

\begin{document}
\begin{frontmatter}


\title{The identification of spatially constrained \\ homogeneous clusters of Covid-19 transmission}

\author[add2]{Roberto Benedetti}
\ead{Benedett@unich.it}
\author[add3]{Federica Piersimoni}
\ead{piersimo@istat.it}
\author[add4]{Giacomo Pignataro}
\ead{giacomo.pignataro@polimi.it}
\author[add1]{Francesco Vidoli} 
\ead{francesco.vidoli@uniroma3.it}

\address[add2]{Department of Economic Studies, G. d'Annunzio, University of Chieti-Pescara, Italy}
\address[add3]{Istat, Directorate for Methodology and Statistical Process Design, Rome, Italy}
\address[add4]{Department of Management, Economics and Industrial Engineering, Politecnico di Milano, and Department of Economics and Business, University of Catania, Italy}
\address[add1]{Department of Political Science, University of Roma Tre, Italy}

\begin{abstract}
The paper introduces an approach to identify a set of spatially constrained homogeneous areas maximally homogeneous in terms of epidemic trends. The proposed hierarchical algorithm is based on the Dynamic Time Warping distances between epidemic time trends where units are constrained by a spatial proximity graph. The paper includes two different applications of this approach to Italy, based on different data (number of positive test and number of differential deaths, with respect to the previous years) and on different observational units (provinces and Labour Market Areas). Both applications, above all the one related to Labour Market Areas, show the existence of well-defined areas, where the dynamics of growth of the infection have been strongly differentiated. The adoption of the same lock-down policy throughout the entire national territory has been therefore sub-optimal, showing once again the urgent need for local data-driven policies.
\end{abstract}

\begin{keyword}
Spatial justice \sep Spatial heterogeneity \sep Spatial clustering \sep Epidemiological Models \sep R(t) \sep Time series distance 
\end{keyword}

\end{frontmatter}

\section{Introduction }
\label{introduzione}

Coronavirus disease 2019 (Covid-19) caused by SARS-CoV-2, is an infectious disease that was first identified  in 2019 in the Hubei province of China, in particular in the city of Wuhan. By the middle of May 2020, the pandemic had spread to 188 countries.

As data on daily contagions and on the reproduction number are progressively improving (China first and Europe afterwards), countries are lifting the lockdown measures, enforced since the beginning of the pandemic. In the discussion on how to design such measures, the crucial policy trade-off between health safety and a quick economic recovery (taking into account that prolonged economic recessions have historically brought forward long-term indirect consequences on health) needs to be considered. The concern about this trade-off is generally requiring a decision on how far to go in lifting the previous restrictions on mobility and contacts, with all the implications on the way economic, social and individual activities can be carried out (\citealp{bonaccorsi2020evidence}). 

However, because of the general non-uniform spread of the contagion within a country, the other relevant policy issue is whether to have a differentiated implementation of the lift of the lockdown restrictions for different geographic areas (what \citealp{Friedman2020} call a zone-based social distancing, or the geographic segmentation hoped for by \citealp{WHO2020b})\footnote{More generally, policies can be differentiated for different groups of individuals, focusing non only on the place where they live, but on other characteristics, like their age or the industry where they work. Our interest, however, is on spatial differentiation. }.  Actually, this was an issue also in the first place, when these restrictions were introduced. 
After a first period in which choices targeted on specific local situations were more frequent (for example the delimitation of restricted circulation zones around an outbreak, i.e. the so-called red areas), governments have increasingly implemented identical rules all over the country, slowing down the economy and trade even in areas where it was probably not necessary. Whenever the issue of having different rules for different areas of a country has been raised, it has been referred to its administrative areas (at different levels: states, regions, municipalities, etc.) as a reflection of the internal distribution of the decision-making powers, relevant for the take-up of the measures. In Italy, for instance, the institutional debate on how differentiated the lifting of lockdown restrictions should be is focused on the differences across the regions, as supported by the settlement of a monitoring system at that territorial level. 

Even in the short time since the outbreak of the Covid-19 pandemic, there is evidence\footnote{The effort of the scientific community has taken advantage from most governments' efforts to provide up-to-date epidemiological data at different levels of resolution, from the national aggregates to the regional and the sub-regional breakdowns, and to make them publicly available. For Italian data see \url{https://github.com/pcm-dpc/COVID-19}.} that the infection followed specific patterns dictated by the territorial proximity, by spatial human mobility (\citealp{Kraemer493, Gatto2020}) and that it was strongly concentrated in some areas. In other words, the pattern of contagion developed upon to a pre-existing spatial organization, which is not necessarily characterized, at a geographic level, by the coincidence with whatever administrative boundaries. Several studies have pushed to consider the differentiation of the virus outbreak areas in terms of relevant characteristics of disease transmission (among others, \citealp{Harris2020,Siegenfeld2020,Zhao2020}) and some have explicitly considered the tool of spatial analysis and its main concepts of spatial dependence and spatial heterogeneity (\citealp{Bourdin2020}). 

The objective of this paper is the identification of geographic areas characterized by different patterns of Covid-19 transmission. This is a problem connected with the partitioning of spatial data with respect to the trends of the epidemic curves. While there are by now several attempts to deal with spatial aspects in the health field, from spatial dependence (\citealp{Baltagi2018}) to spatial heterogeneity (\citealp{Baltagi2017,Auteri2019}), a specific, systematic and complete approach to such a grouping problem is not yet available and it is still the focus of scientific debate (\citealp{Zhou2020}). \\
We introduce a methodology, which develops in three subsequent steps. First, for each geographic area, considered as the elementary observational unit of the analysis, we measure the extent of the virus transmission, which represents the basic information for the sort of decision-making problem discussed earlier. As it is standard, we use an estimate of the time dependent reproduction number $R(t)$ time series, built on the row data, which directly (the number of cases) or indirectly (the differential number of deaths with respect to homogenous time period of previous years) measure the daily incidence of the disease.  Second, we try to capture the extent at which the $R(t)$ trends of the different observational units are similar to each other, by using the \emph{Dynamic Time Warping} algorithm, which provides a measure of the distance between the time series of the $R(t)$ indicator. Third, we employ the \emph{Skater} algorithm to combine information about the spatial proximity of the observational units and the difference of their time series of the $R(t)$ indicator, so as to identify spatial clusters of the units -- maximally homogenous within and heterogenous between. The application to the Italian case allows to provide a more precise geographic picture of the differences in the pattern of Covid-19 transmission and, therefore, can be regarded a suitable information basis for an appropriate geographic modulation of the policies aimed at containing the virus transmission while avoiding useless disruptions of the economic and social activities. 

The paper is structured as follows. In Section \ref{metho}, we present the empirical methodology for the identification of homogenous geographic clusters of Covid-19 transmission. In Section \ref{applic}, this methodology is applied to Italy: while in Section \ref{subapplic1}, our observational units are the Italian provinces and the $R(t)$ time series are estimated on the basis of the daily provincial number of cases, in Section \ref{subapplic2} we switch to different observational units, the Labor Market Areas (LMAs), and to data on the municipal variation of the daily deaths (\citealp{Modi2020}).  Finally, in Section \ref{fine}, the main results of the paper are discussed and some important issues related to policy implications and decisions are identified.


\section{A methodological approach for estimating spatially constrained zones}
\label{metho}

The proposed analytical framework runs essentially through three main steps. In the first one, the Real-Time Reproduction Number $R(t)$ trends (\citealp{Wallinga2004}) are estimated separately for each unit of analysis. Such time trends are, then, compared all each other by means of the Dynamic Time Warping algorithm, so as to obtain a distance matrix with the aim of estimating a synthetic measure of the difference between time series. Finally, the estimated distance matrix together with information on the proximity among units had been used to identify clusters of neighbouring areas maximally homogeneous in terms of time trends, through a redesigned version of the \emph{Skater} (Spatial K'luster Analysis by TreeEdgeRemoval, \citealp{Assuncao2006}) algorithm.

\subsection{Real-Time Reproduction Number and the Dynamic Time Warping distance}
\label{submetho1}

The key epidemiologic variable that characterizes the potential transmission of a disease is the basic time dependent reproduction number, $R(t)$, which is defined as the expected number of secondary cases produced by a typical primary case in an entirely susceptible population (\citealp{Wallinga2004}). The higher the value of this indicator, the higher the risk of spreading the epidemic. Since the beginning of this epidemic, the World Health Organization (WHO) and numerous research institutes around the world have released estimates of this parameter. \\
$R(t)$ is a function of \emph{(i)} the probability of transmission by single contact between an infected and a susceptible person, \emph{(ii)} the number of contacts of the infected person and \emph{(iii)} the duration of the infectivity; reducing at least one of the three growth factors can reduce this value and therefore be able to control, or at least delay, the spread of the pathogen to other people. The probability of transmission and the duration of infectivity (without a vaccine or a treatment that reduces viraemia) are not modifiable at this stage but, the immediate diagnosis and identification of the infected person, or of the potentially infected person, and the possibility of reducing his/her contacts with other people would allow a reduction of $R(t)$. To stop an epidemic, $R(t)$ needs to be persistently reduced to a level below 1. The estimate of $R(t)$ is quite simple if we have information on who has infected whom, in these cases it is possible to build an infection network, in which connections are active if one person has infected the other. Estimating $R(t)$ simply involves counting the number of secondary infections for each unit (\citealp{Wallinga2004}).

Often, estimation is a much more complicated affair, because only the epidemic curve is observed and there is no information on who has been infected by whom. However, in most cases, the approximation of $R(t)$ is used assuming a theoretical trend in the number of cases over time and adapting to observed data this specific model which summarizes the assumptions about the epidemiology of the disease (\citealp{Gani2001,Riley2003}).

From the time series of the relevant observation variable (either the number of positives or the differential number of deaths), a maximum likelihood (ML) estimate is obtained, opting for the hypothesis of time dependence of the parameter $R0$, namely $R(t)$, as suggested by \cite{Wallinga2004}. In fact, it cannot be considered stationary over time, both for stay-at-home measures and because the cognitive objective is to see its evolution over time until it can be considered sufficiently low to allow the government to reopen economic activities even if with more or less binding rules.

The similarities between $R(t)$ trends among different spatial units have been assessed using dynamic time warping, a widely used technique whose rationale is to locally stretch or compress two time series in order to make one resemble the other as much as possible.  In such a way, it provides a measure of distance, insensitive to local compression, stretching (namely ``warping'') the relative curves and optimally deforming one of the two input series onto the other (\citealp{Giorgino2009}). The distance is thus computed, after stretching, by summing the distances of individual aligned elements.

This technique has long been known in the speech recognition community. It allows a non-linear mapping from one time series to another minimizing the distance between the two. DTW was introduced to the Data Mining community as a utility for various activities for time series issues including classification, clustering and anomaly detection (\citealp{Montero2014,Pree2014}). The technique has spread rapidly and has been applied, as well as in field of speech recognition, to a great deal of problems in various disciplines including handwriting and online signature matching, finger print verification, pattern and shape recognition, computer vision and animation, surveillance, protein sequence alignment, chemical engineering, music and signal processing (\citealp{Yadav2018}).

\subsection{Hierarchical spatially constrained clustering algorithm based on time-series distances}
\label{submetho2}

The redesigned procedure\footnote{The relative R functions - derived from the \emph{spdep} package functions - are available from the authors upon request.} of the \emph{Skater} (\citealp{Assuncao2006}) algorithm aims at identifying $k$ not overlapping clusters of units that are geographically proximate and as much as possible homogenous in terms of the measure of time-trend distance. More in detail, the \emph{Skater} procedure can be described as a $k$-means clustering procedure in which the units are belonging to a proximity graph: each observation, thereby, belongs to the cluster with the nearest mean (measured in terms of distance) in order to partition $n$ neighbouring observations into $k$ clusters.

Conceptually, the algorithm can be split into two main steps: \emph{(i)} the identification of geography and distances among units, and \emph{(ii)} a second phase in which the effective spatially constrained clustering algorithm is implemented. In the first step, the units are first represented as a full neighbourhood graph and then this complete network is simplified according to the Minimum Spanning Tree algorithm (MST, \citealp{Pettie2000}).
Starting from this simplified representation of the neighbourhood, the purpose of the second phase is to identify spatial clusters - maximally homogeneous within and heterogeneous between - in terms of time-trend distances. 

The proposed redesigned procedure for time series differs from the standard \emph{Skater} algorithm essentially in the objective function to be maximized during the cluster search phase: in the original algorithm, the different sub-graphs are compared in terms of intra-cluster square deviation for a set of variables, while in this case comparison faces in terms of dynamic time warping distance between trends. More formally, in a generic step, unit $k$ is included in cluster $A$ if the average Dynamic Time Warping ($dtw$) distance between its trend and that of the other units belonging to cluster $A$ ($dtw_{(k,A)}$) is minimal compared to other units $q$ other than $k$ not yet belonging to $A$ and spatially contiguous to $A$. In each step:
\begin{equation}
k \in A, \text{ if } dtw_{(k,A)} < dtw_{(q,A)} \forall q \neq k \text{ \& } k,q \text{ contiguous to } A
\end{equation}
where
\begin{equation}
dtw_{(k,A)} = E(dtw_{i,j}), \forall i,j \in A \cup k
\end{equation}

Please note how the complexity of the algorithm, written in such schematic terms, can grow very fast as the units under analysis grow. \cite{Assuncao2006} write that "\emph{the exhaustive comparison of all possible values of the objective function is expensive computationally [and it] leads to a combinational explosion}". In order to practically overcome this drawback and reduce the search complexity,  \cite{Assuncao2006} suggests to look for the edges no longer among all the possible nodes, but only among those already calculated.

For other technical details, please refer to \cite{Assuncao2006}.


\section{The identification of spatial clusters of Covid-19 transmission in Italy}
\label{applic}

We carry out two different attempts of applying the methodology outlined in Section \ref{metho} for the identification of homogenous geographic areas in terms of epidemic trend, to Italy. The difference between the two applications is related to the spatial observational units of our analysis and to the nature of the data used for the estimation of the time trends of the $R(t)$ indicator for each unit. In Section \ref{subapplic1}, we consider as the basic geographic unit of analysis the Italian provinces and we use (provincial) daily data on the number of Covid-19 cases; in Section \ref{subapplic2}, we move to focus the application on different geographical units, namely LMAs, and we use smooth daily data at week level for the municipal variation of deaths (with respect to the analogous time period of the previous years), aggregated at the LMA level.

\subsection{Estimation of clusters using provincial Covid-19 diseases data}
\label{subapplic1}

We use the daily data on the total number of Covid-19 positive cases, published daily by the Civil Protection Department, for each of the 101 Italian provinces\footnote{Source: \url{https://github.com/pcm-dpc/COVID-19}, last analysis update: May 15, 2020. The Civil Protection Department time series starts on 24$^{th}$ February, 2020.}. It is important to notice that the data reported by the Civil Protection Department actually refer to the number of positive (to Covid-19) tests, without differentiating between diagnostic and control tests. Therefore, it cannot be considered a measure of the daily incidence of the disease. 

\begin{figure}[!htbp]
\centering
\includegraphics[width=.60\columnwidth]{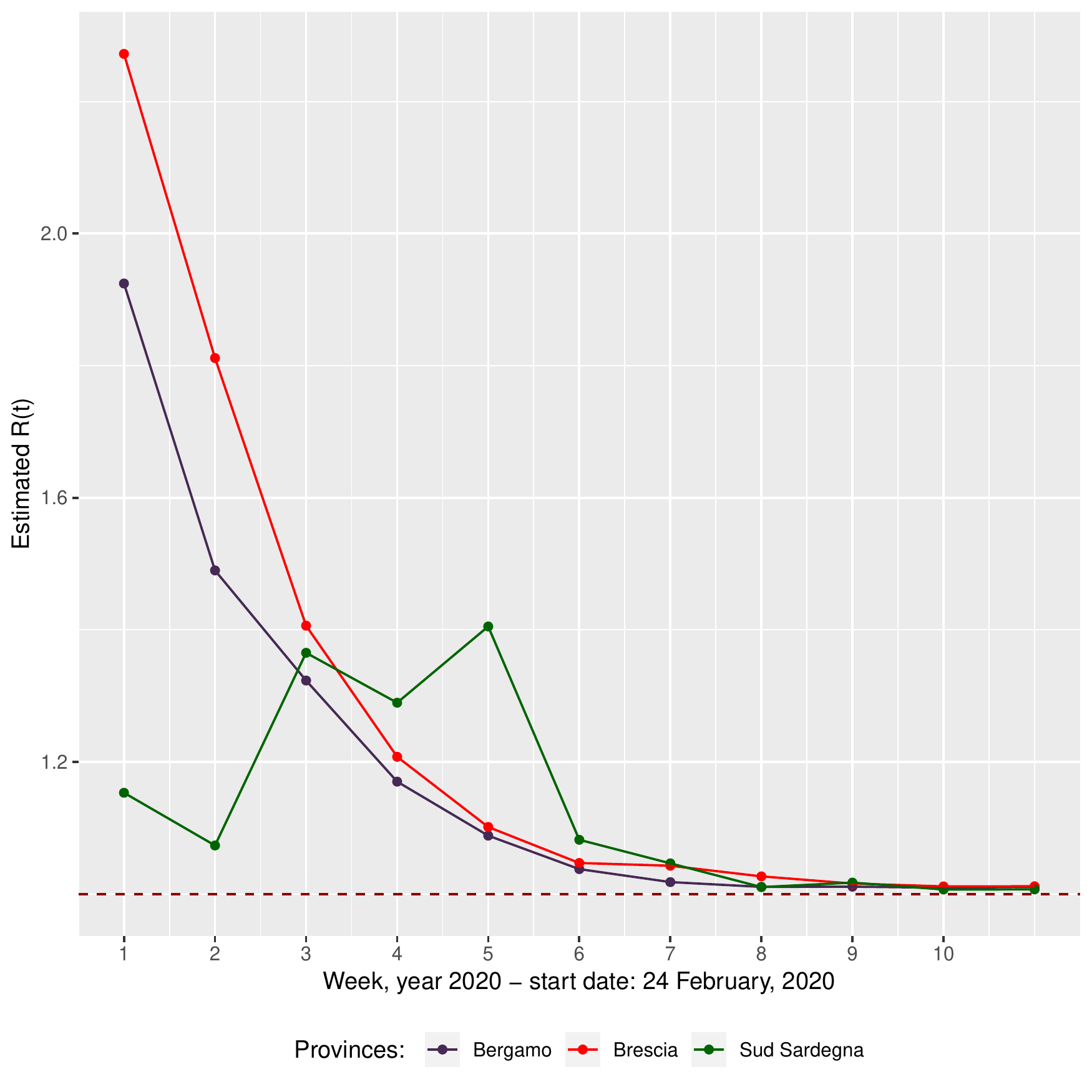}
\caption{Clusters estimated in terms of the epidemic trend - Italy}
\label{PROV_PGM2_R0trend}
\end{figure}

The first step of the analysis has involved the estimation of the time dependent reproduction number $R(t)$ according to \cite{Wallinga2004} specification distinctly for each Italian province. The estimated real-time reproduction number trend obtained for each day has been subsequently smooth over weeks so as to obtain more robust estimates.
The basic idea is precisely to compare estimated epidemic trends for each province; Figure \ref{PROV_PGM2_R0trend} shows, for example, the estimated trends for three provinces: Bergamo and Brescia, which show very similar trends, and  Sud Sardegna, which has experienced a lower growth trend (also in absolute terms).

\begin{figure}[!htbp]
\centering
\includegraphics[width=.60\columnwidth]{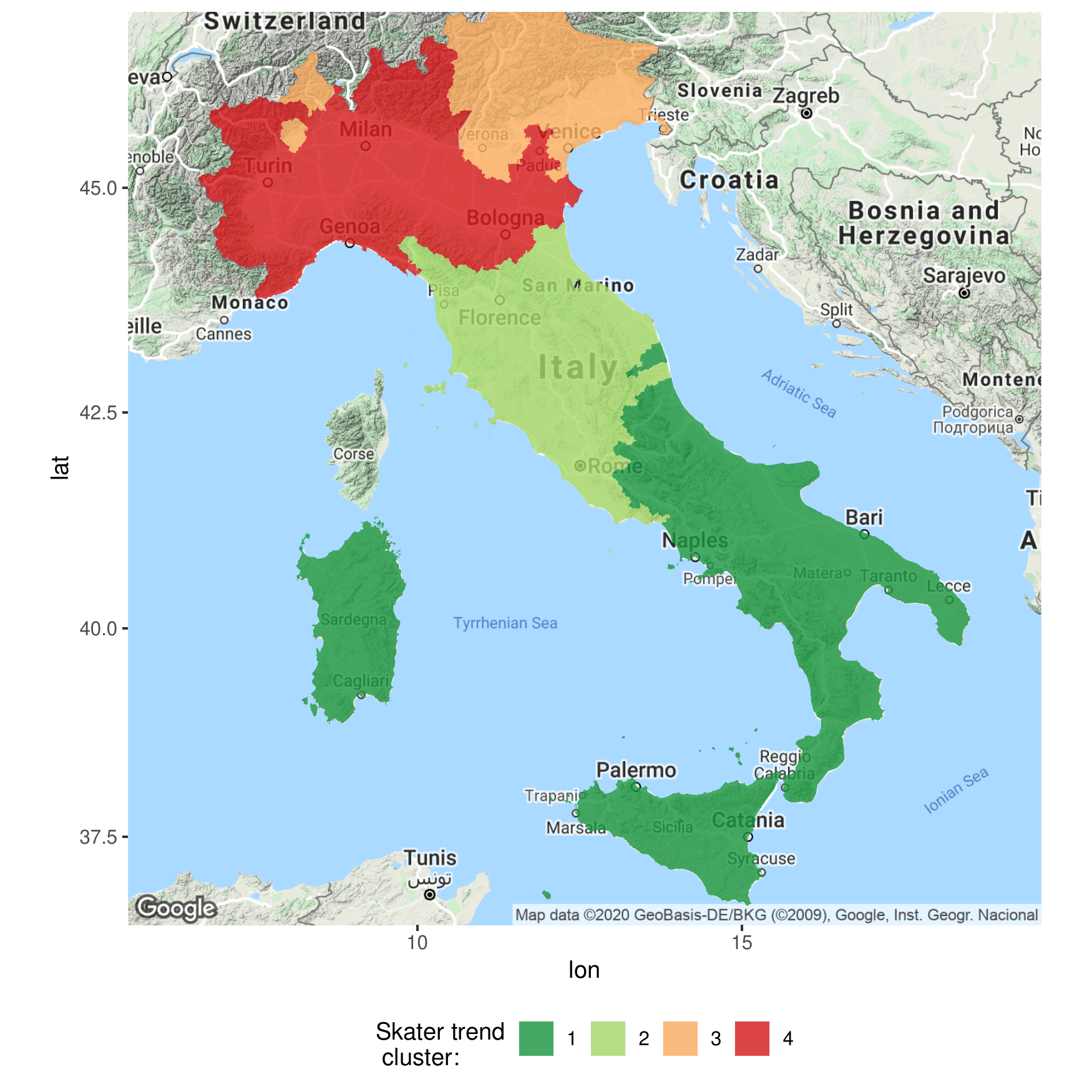}
\caption{Clusters estimated in terms of the epidemic trend  - Italy}
\label{PROV_PGM5_Italia4}
\end{figure}

In the second step, using the estimated provincial $R(t)$ trend, the Dynamic Time Warp distance matrix has been calculated. To overcome the potential misalignments among time series due to the differentiated impact of the epidemic over space, the optimal alignments between time series have been computed for every comparison (\citealp{Giorgino2009}).\\
Finally, starting from the provincial centroids, the minimum spanning neighbourhood graph has been calculated (see Figure \ref{PROV_PGM4_vicinato}) in order to apply the \emph{Skater} procedure. . Four homogeneous zones, estimated in terms of homogeneity with respect to epidemic trends and geographic proximity of the provinces (see Figure \ref{PROV_PGM5_Italia4}), emerge. They show a clear differentiation between Northern Italy, with a split between North-West and North- East, Central Italy including Tuscany and part of Emilia-Romagna, and the South where the epidemic peak has been averted.

\begin{figure}[!htbp]
\centering
\includegraphics[width=.60\columnwidth]{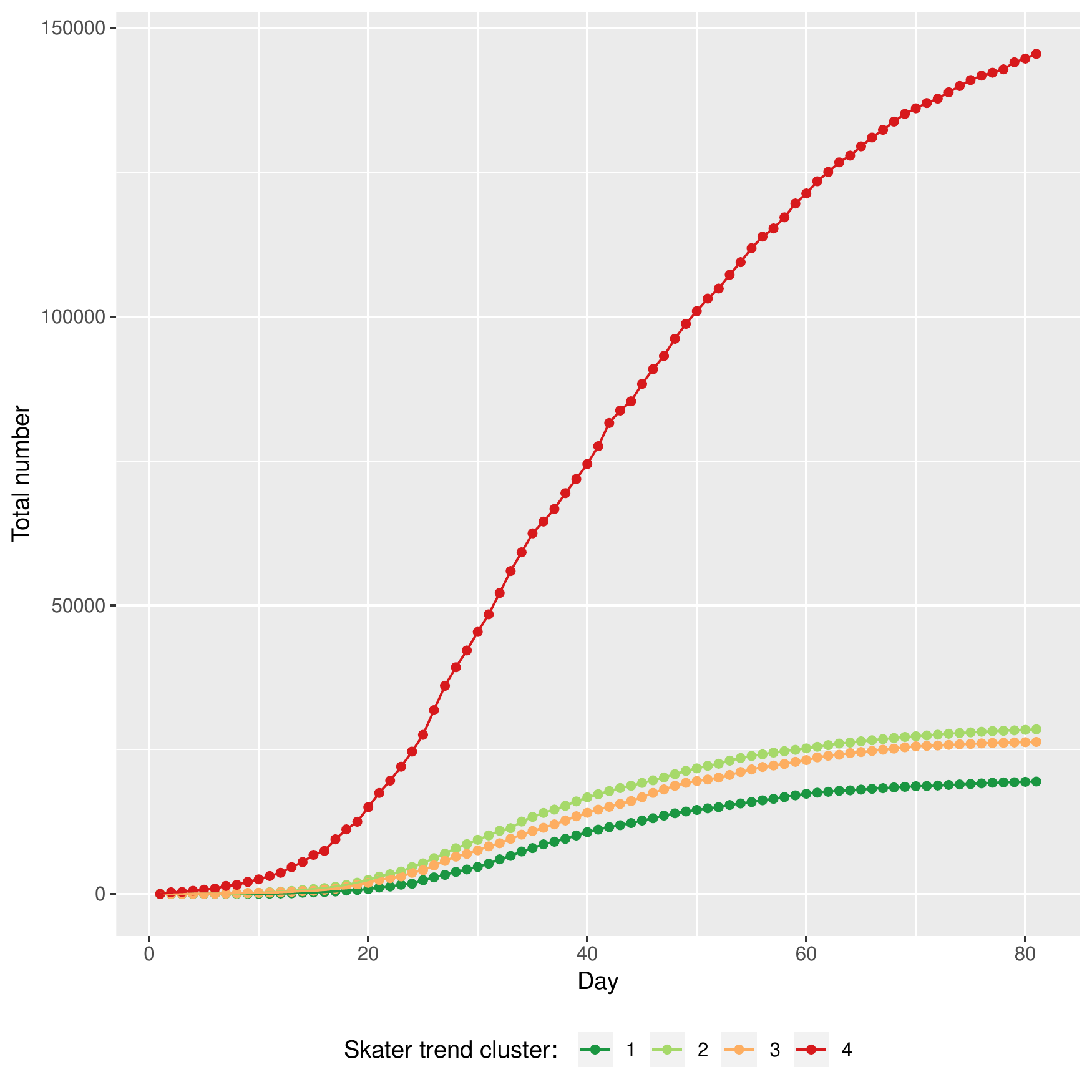}
\caption{Total number of positive tests by spatial cluster, as estimated in terms of epidemic trend - provincial level}
\label{PROV_PGM6_trend_num}
\end{figure}

The great differentiation between homogeneous zones, both in terms of quantity and growth rate, of official infected people can be valued in Figure \ref{PROV_PGM6_trend_num}, where there is a clear division between geographical areas, especially between cluster 4 (North-West) and the rest of Italy. 


\subsection{Estimation of clusters using Municipal deaths data}
\label{subapplic2}

The clusterization carried out in Section 3.1 raises two key questions. The first one is related to the homogeneity of the data on the daily number of positive tests, both in terms of its distribution between diagnostic and control tests and of the intensity of testing in each province, with respect to population. One of the main implications for the application of our methodology is whether the relationship, over time, between tested positive and actually infected people can be considered constant in space. The answer is negative\footnote{Figure \ref{SSL_PGM0_timeseries2} shows how the ratio of the number of positive tests and mortality trend changes clearly across provinces.}, since, due to the highly decentralized nature of the Italian National Health Service, the provincial data on the number of positive tests reflect the testing strategy (and its intensity in terms of population tested) adopted by each. For this reason, and because of the possible problems of underestimation\footnote{The potential underestimation has been underlined by several official studies (see for example the study of the English ONS at \url{https://www.ft.com/content/67e6a4ee-3d05-43bc-ba03-e239799fa6ab}, or that of INPS for Italy at \url{https://www. inps.it/nuovoportaleinps/default.aspx?itemdir=53705}), which estimate the deaths due to Covid-19 in about the double the official rate.}  of the actual infection rate, many authors (see \emph{e.g.} \citealp{Modi2020,Ghislandi2020}) have suggested using mortality data differentials between the year 2020 and previous years.

The second question arises from the observation of the clusterization of provincial units, which has a limited informational content, since we get only four clusters. The number of clusters is constrained by the limited number of spatial units to be aggregated (101 overall). To improve the informational content of our clusters, therefore, we need to look for a more detailed geographic delimitation of the spatial units to be aggregated in clusters and, above all, they should be correlated to some pre-existing spatial organization that is relevant for the transmission pattern of the disease, thus avoiding a spatial delimitation that simply refers to administrative boundaries. More precisely, we look for geographic areas that can be regarded as homogenous in terms of functional mobility\footnote{\cite{OECD2002} (p.11) defines a functional region as ``\emph{a territorial unit resulting from the organisation of social and economic relations in that its boundaries do not reflect geographical particularities or historical events. It is thus a functional sub-division of territories}''.}. 

Similarly to existing literature (see \emph{e.g.} \citealp{MartinezBernabeu2012,Chakraborty2013,Franconi2017}), we use the municipal functional aggregation called Labor market areas (LMAs), provided by the Italian National Institute for Statistics (ISTAT); more specifically, LMAs are ``\emph{sub-regional geographical areas where the bulk of the labour force lives and works, and where establishments can find the largest amount of the labour force necessary to occupy the offered jobs}''. \cite{Monras2020} has stressed the relevance of labour mobility for the trade-off between health safety and economic recovery: \emph{``we need labour immobility to fight the virus, but we also need people to move for the economy not to collapse}''. Most of the mobility, which affects the speed of the virus transmission, is ``\emph{concentrated around work and home locations}'' and, therefore, ``\emph{commuting zones may offer a way to think about designing more targeted policies to halt the spread of Covid-19}'' (\citealp{Monras2020}). \\
Altogether, ISTAT classifies 610 LMAs based on commuting data stemming from the 15$^{th}$Population Census using an allocation process shared at European level. It surely represents a more refined partitioning of the country than the one represented by the provincial partition. The information provided by ISTAT allows to identify the municipalities aggregated in each LMA and, therefore, we are able to link the mortality data, collected at the municipal level, with the LMA partition and to aggregate these data at the LMA level. 

\begin{figure}[!htbp]
\centering
\includegraphics[width=.60\columnwidth]{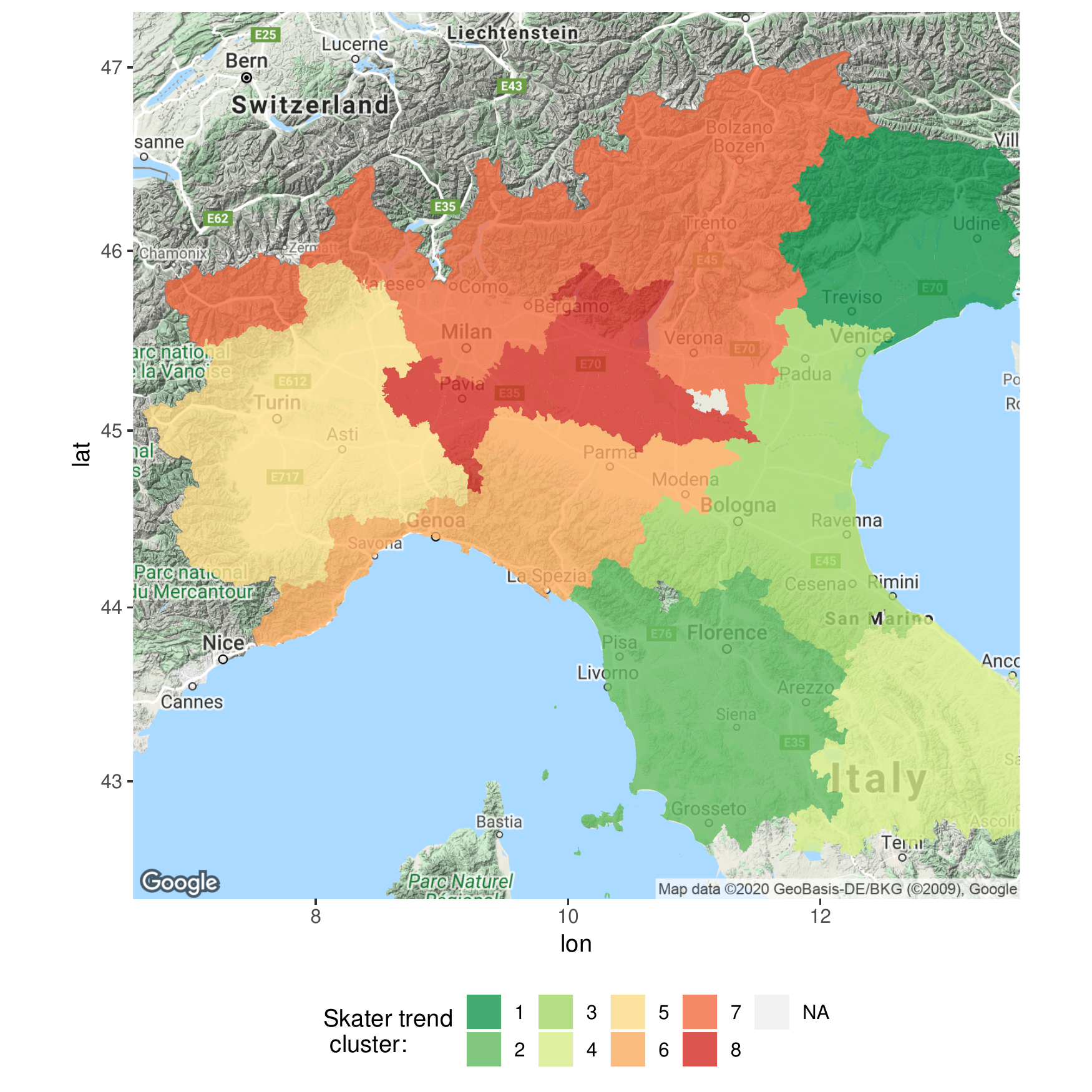}
\caption{Clusters estimated in terms of the mortality difference trend - Northern Italy}
\label{SSL_PGM5_Italia4}
\end{figure}

\begin{figure}[!htbp]
\centering
\includegraphics[width=.60\columnwidth]{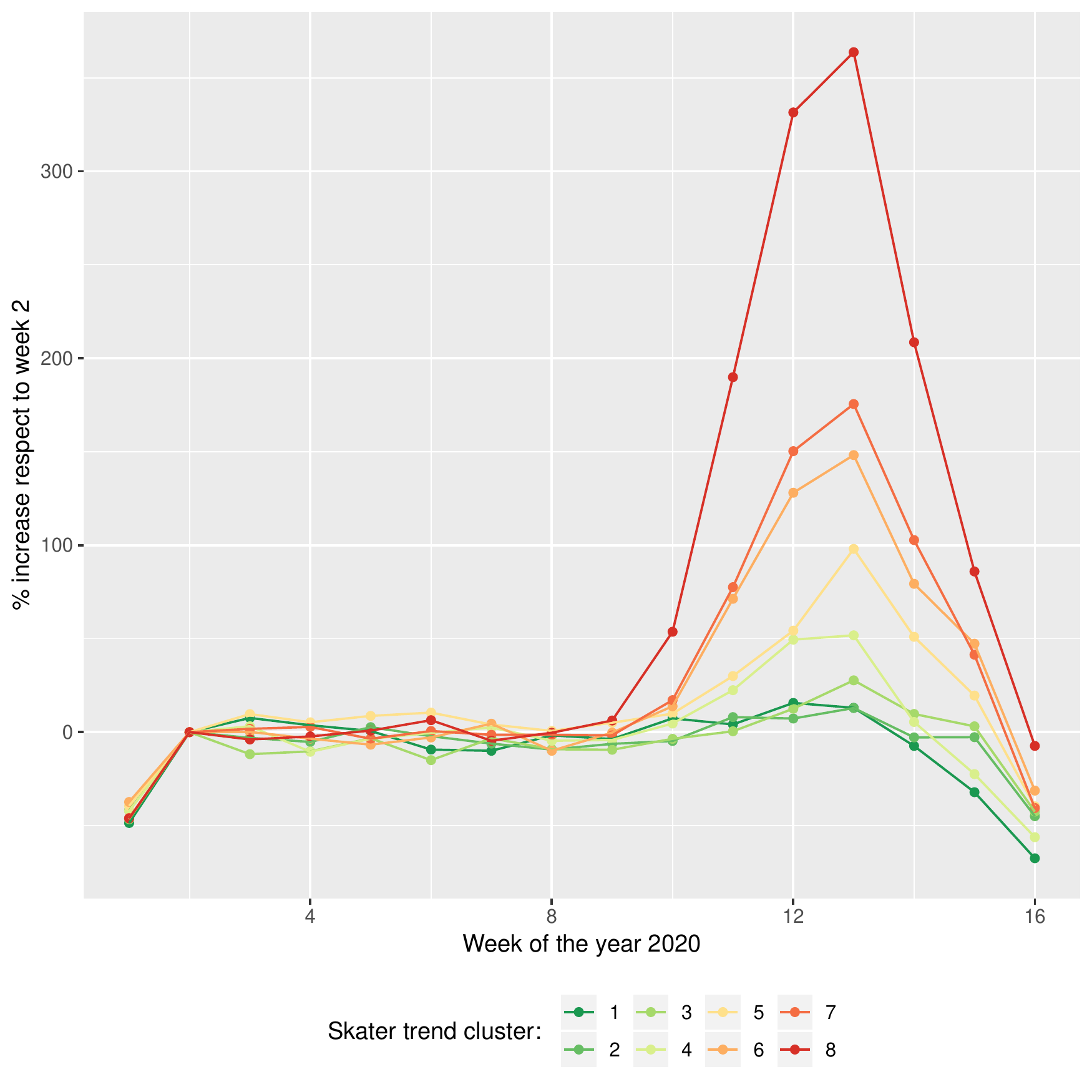}
\caption{Percentage increase in deaths compared to the second week of the year by spatial cluster - Northern Italy}
\label{PGM6_R0_trend_num}
\end{figure}

Against this background, data on the difference in mortality between the year 2020 and the average for the years 2015-2019 at municipal level have been used\footnote{Source: \url{https://www.istat.it/it/archivio/240401}, last analysis update: May 15, 2020. Figure \ref{SSL_PGM0_timeseries} shows both the clear difference between the years 2015-2020 starting approximately from the beginning of March, and the gradual unreliability of the statistical data from 1$^{th}$ April due to the delay in the update.}. Starting from the mortality data, aggregated at LMA level, the single real-time reproduction number trend $R(t)$, the matrix of distances between the estimated trends and, finally, the homogeneous areas in terms of trends of virus transmission have been estimated -- similarly to what done in Section \ref{subapplic1}, for the estimation of clusters based on the provincial level. Since the burden of the Covid-19 disease (in terms of cases and deaths, as well as of demand for hospital services) has been concentrated in the Northern part of Italy, as confirmed by our clusterization results in Section \ref{subapplic1}, we decided to focus on  the regions of Northern Italy, just to check whether a different choice of the observational units (the LMAs) can provide a more refined information of the geographic differences in terms of virus transmission. Our methodology, this time, is therefore applied to the sub-sample of LMAs including Municipalities of all the Northern Italy regions. Figure \ref{SSL_PGM5_Italia4} actually shows that while in the previous clusterization the provinces of these regions were part of just three clusters (see Figure \ref{PROV_PGM5_Italia4}), their LMAs are now differentiated in 8 clusters. In terms of the severity of the $R(t)$ time trend, the most badly hit area no longer covers the entire North west of the country, but it is limited to cluster 8. Figure \ref{PGM6_R0_trend_num} shows values of the growth rate of the differential of mortality, with respect to the second week of the year, for cluster 8, up to more than 300\%. This cluster is made up of LMAs aggregating almost entirely municipalities of the Lombardy region (only 14 municipalities, out of 789, are part of other regions – Piemonte, Veneto and Emilia Romagna). Moreover, Lombardy is not to be considered a uniform geographic area, in terms of the $R(t)$ time trend, but its territory is distributed between two clusters, 7 and 8. About 45\% of its municipalities, representing 35\% of its population, are part of the most badly hit area, cluster 8, while the rest is part of cluster 7, which shows a significantly different dynamic pattern of the contagion (see Figure \ref{PGM6_R0_trend_num}).  Clusters 5, 6 and 7, although with differentiated growths rates, lower than the ones characterizing cluster 8, show a separate trend from the rest of the estimated areas.

\subsection{Discussion of results}
\label{discuss}

The results presented in the two previous sections \ref{subapplic1} and \ref{subapplic2} represent an attempt of identifying spatially constrained homogenous zones, on the basis of the estimated values of $R(t)$.\\
First of all, we would like to point out the qualitative differences, in the nature of the information arising from the zone mapping, between the two different applications of the methodology developed in section \ref{metho}, and also with respect to the current monitoring system of the epidemic, implemented by the Italian Ministry of Health, whose indicators, including $R(t)$, are estimated on a regional basis\footnote{The monitoring system, implemented since the beginning of May 2020, is the outcome of a cooperation between the central government and the regional governments. It is meant to be an information tool for the design of the reopening measures, based on epidemiological data and on the measurement of the regional capacity of early response. Indicators include a measure of the weekly incidence of the disease, an estimate of $R(t)$, the weekly trend of the number of positives (whether declining or not), a categorical evaluation of the change in transmission and on its impact on the regional healthcare system, a measure of the resilience of the community healthcare services. }. \\
The regionalization of the information provided by the official monitoring system is basically motivated by institutional reasons. Regional governments, in Italy, are granted, by Constitution, autonomous powers for the provision of healthcare services (within the framework of nationally uniform general principles about the fundamental architecture of the healthcare system) and, therefore, bear the responsibility for all the measures, which involve the utilization of their healthcare systems. However, the regional basis of the monitoring system is not necessarily the relevant one for the design of the other measures for the management of the different stages of the epidemic. It is in fact disputed that, on the basis of Italian Constitution, regional governments should have a say on community mitigation strategies. Moreover, the latter should anyway be consistent with the geographic patterns of the virus transmission, so as to balance health and economic safety. 

Our estimation of $R(t)$, therefore, tries to overcome the limitations arising from sticking to administrative boundaries that are only partially relevant from an institutional point of view and risk to be inconsistent with the epidemic geographic pattern. This inconsistency is only partially overcome in our first attempt of identifying homogenous clusters on the basis of the provincial data on the number of contagions (section \ref{subapplic1}). The clusters, because of its limited number, overlap the regional boundaries while, at the same time, there are regions whose territory is shared by different clusters. However, this clusterization has severe limitations of the information it provides, arising from the limited number of clusters, constrained by the relatively small number of observation units (101 provinces all over the country), and from the provincial source of the data on the number of positive tests, which is not necessarily related to the geographic patterns of the virus transmission. 

The application in section \ref{subapplic2} is, instead, based on smaller and more numerous observation units but, above all, on a pre-existing spatial organization, which is surely relevant for the transmission pattern of the virus. As already noticed, the LMAs are identified through the analysis of the work commuting patterns and, therefore, are related to an essential component of daily mobility, which is obviously considered as one of the main drivers of a virus transmission. In such a way, when we aggregate LMAs in a cluster, by spatial proximity and by homogeneity of the $R(t)$ time trend (as measured according to the \emph{Dynamic Time Warping} algorithm), we take in municipalities, which may have different values of $R(t)$ with respect to the one for the entire LMA, but are however connected to the rest of this area by a work mobility pattern. \\
The differentiation of the geographic areas we get is undoubtedly finer than the one based on the regional values of $R(t)$ and able to get a discrimination of geographic areas within a region. Our focus on Northern Italy allows to consider the situation of those regions, which were badly hit by Covid-19 -- Piemonte, Lombardy, Veneto, Emilia Romagna and Tuscany, and are usually regarded as a sort of ``uniform'' block. 

The clusterization based on LMAs, however, show a very differentiated picture. As far as Piemonte and Tuscany are concerned, each of these two regions appears to be quite homogenous in terms of our estimate of $R(t)$. Each region is almost self-contained in one cluster, respectively 5 and 2, with the few exceptions of those municipalities aggregated in LMAs overlapping with other regions. As Figure \ref{PGM6_R0_trend_num} reveals, they show a different pattern with respect to the other regions and clusters, especially Tuscany (cluster 2). Lombardy, as already noticed, is substantially covered by two clusters, 7 and 8, the ones with the most severe estimated values of $R(t)$, even if with a substantially differentiated pattern between the two. The situation of Veneto and Emilia Romagna is much more articulated. The municipalities of Veneto belong to three different clusters (1, 3 and 7, with only 9 over 574 municipalities aggregated to cluster 8), almost evenly distributed among them. While there is a part of Veneto which shows a severity of transmission of the virus homogenous to part of Lombardy (cluster 7), two thirds of the region belong to clusters with substantially weaker values of $R(t)$. Emilia Romagna is covered by two clusters, 3 and 6, with relevant differences from each other. 

The main advantage of this picture and, more generally, of our methodology is the possibility of linking the community mitigations strategies as well as their uplift not to mere administrative areas (\emph{e.g.} regions or municipalities), but to clusters whose components (the LMAs) are homogenous with the other ones belonging to the same cluster in terms of the virus transmission rate, but are also homogenous inside in terms of work commuting habits. It should guarantee a better trade-off of the policy intervention between the goal of health safety (and of preventing unsustainability of the demand for healthcare services like intensive care) and the need of avoiding unnecessary disruptions of economic and social activity (not to speak of the individual liberties). 

A refined geographic identification of the different areas is relevant not only for the short-term social, economic and health consequences of the different policy interventions but also looking at the medium- and long-term effects. Considering LMAs as the elementary unit of any zoning, for instance, may avoid costly breaks of existing economic networks, as it would be the case if they happen to be located on the boundaries of different regions and, consequently, they could be jeopardized by different regional lockdown/reopening policies. Of course, this sort of ``granular'' approach is not without problems, impinging on the ``clash'' between the spatial organization of a geographic area and the current institutional arrangements related to that area. When the geographic areas, regarded as homogenous from the point of view of the strength of the virus transmission, overlap with different jurisdictions, a problem of governance of the different interventions arise. Unless the power to intervene is already centralized or centralization can be enforced without constitutional breaks, the release of policy measures that should be implemented in areas, which are part of different jurisdictions (\emph{e.g.} different regions or different municipalities), requires coordination. In consideration of the political nature of the (local) governments that need to be involved, coordination can be very costly, above all in terms of time for reaching an agreement, while some of these interventions (like lockdowns) have to be decided very quickly to be highly effective.  The coordination problems are likely to be more relevant the more fragmented is the distribution of the relevant decision-making powers among different jurisdictions (at different levels), as it happens in Italy. As for the specific problems related to the provision of healthcare services, in connection with the treatment of the consequences of epidemic, if the different areas included in the same LMA are part of different regions, it is well possible that the supply of services within that LMA will not be homogenous, in quantity and quality. A potential consequence is that inefficiencies of providers in some areas of an LMA may slow down the reduction of contagions and, consequently, may delay the ``exit'' of the all areas of that LMA from a lockdown, as well as a quick exit from a lockdown may not be safe for the areas of an LMA, which have delays in reducing contagions and may, eventually, result in a policy reversal. 

The issue of the potential asymmetry between the administrative boundaries of the government and of the management of healthcare services and the spatial differentiation of the health needs and demand, revealed by our analysis on the Covid-19 epidemic, is probably more general. As shown, for instance, by \cite{Auteri2019}, who deal with the issue of defining spatial regimes for the estimation of the efficiency of hospital services in Italy, clusters differentiated by the relevant production frontier generally overlap the different regions. Even if we are aware that the definition of the boundaries of jurisdictions is not a short term decision and, however, it has to correspond to a wider spectrum of interests than the ones relative to the provision of a specific service, there is a need to look for institutional coordination mechanisms, flexible enough for guaranteeing a more homogenous policy intervention in specific circumstances, like the ones faced within the current epidemic, as also recommended by \cite{WHO2020a,WHO2020b}.


\section{Final remarks}
\label{fine}

The Covid-19 crisis has found most countries unprepared to offer an effective reaction, both in terms of the supply of healthcare services for treating people infected by the virus, and of the containment of the virus transmission. The main strategy, therefore, has been in terms of extended and massive lockdown of entire countries, with unprecedented negative consequences for their economic and social environment. A fine-tuned policy intervention requires selective measures, which, above all, need to be targeted to the specific outbreak areas. This approach has to be based on appropriate and refined information about the geographic distribution of the virus transmission pattern. 

In this paper, we have outlined a methodological approach, based on a spatially constrained clustering algorithm with the aim of identifying homogeneous areas in terms of epidemic growth. The results show how data of different nature (epidemic and mortality data) and different observational units (provinces and LMA) can provide different pictures of the spatial distribution of the disease. 

The proposed methodological approach is not intended to be an ex post policy evaluation tool, but, given that the population may be vulnerable to new outbreaks in the medium term, is to be intended as a tool for design well-defined potential new areas in terms of epidemic growth. Finally, the proposed framework can be used more extensively in future economic impact assessments, highlighting heterogeneous space-time patterns of regional resilience and showing how the epidemic has changed our lives in social, labour and economic fields.

\newpage
\appendix
\section{ }
\label{allegato1}

\begin{figure}[!htbp]
\centering
\includegraphics[width=.70\columnwidth]{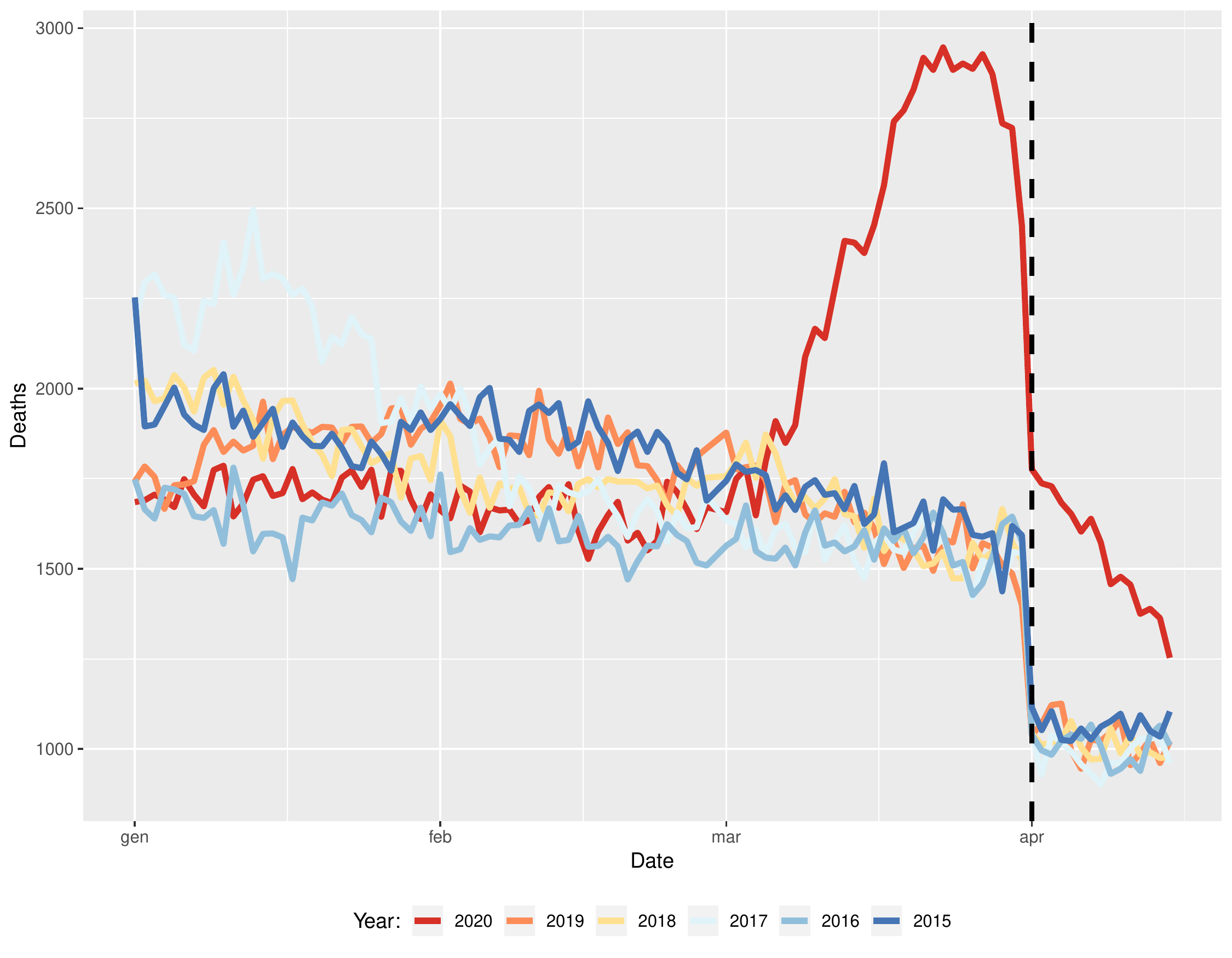}
\caption{Daily mortality per year - years 2015-2020}
\label{SSL_PGM0_timeseries}
\end{figure}

\begin{figure}[!htbp]
\centering
\includegraphics[width=.70\columnwidth]{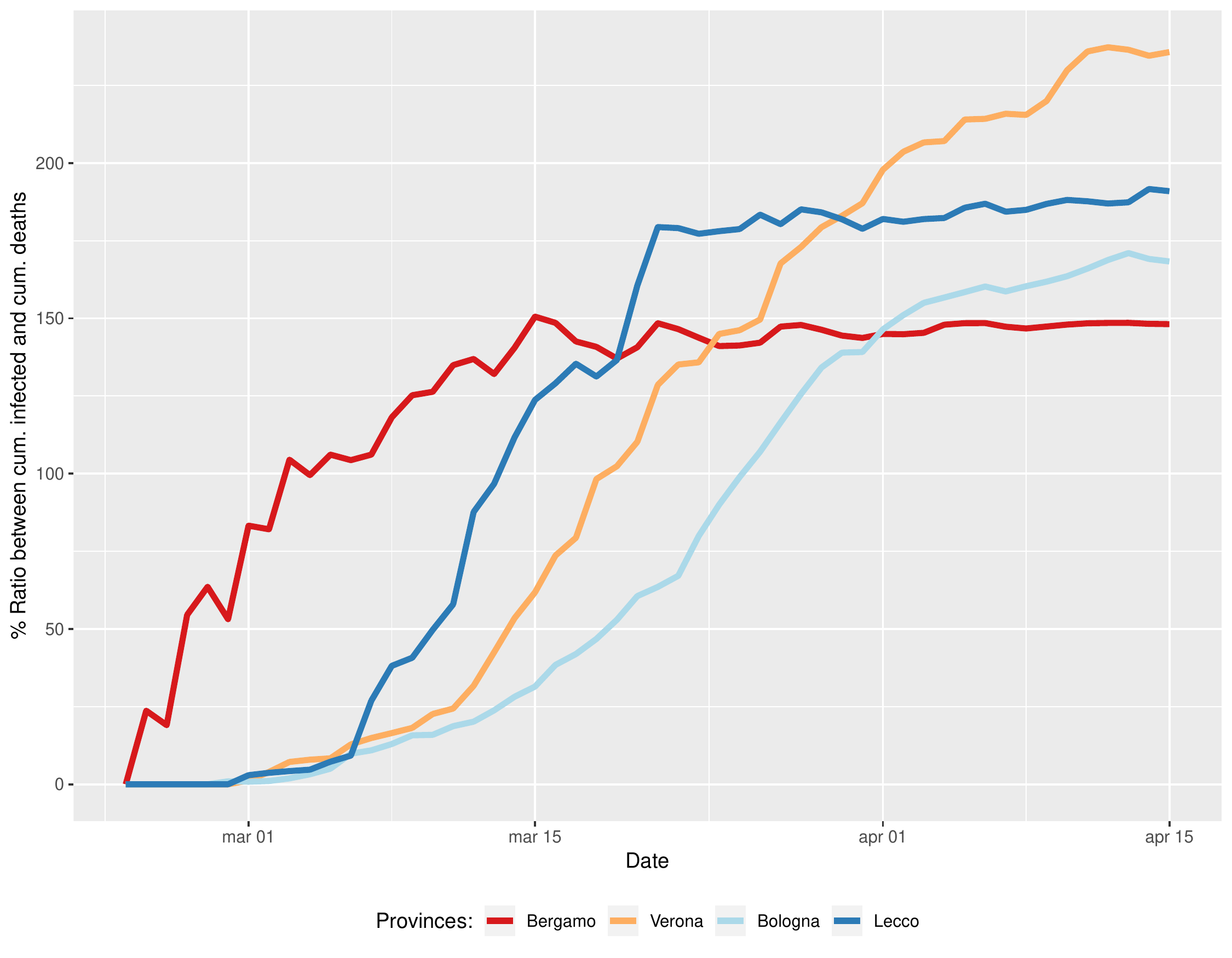}
\caption{Percentage ratio between cumulate infected and cumulate deaths by date - year 2020}
\label{SSL_PGM0_timeseries2}
\end{figure}

\begin{figure}[!htbp]
\centering
\includegraphics[width=.60\columnwidth]{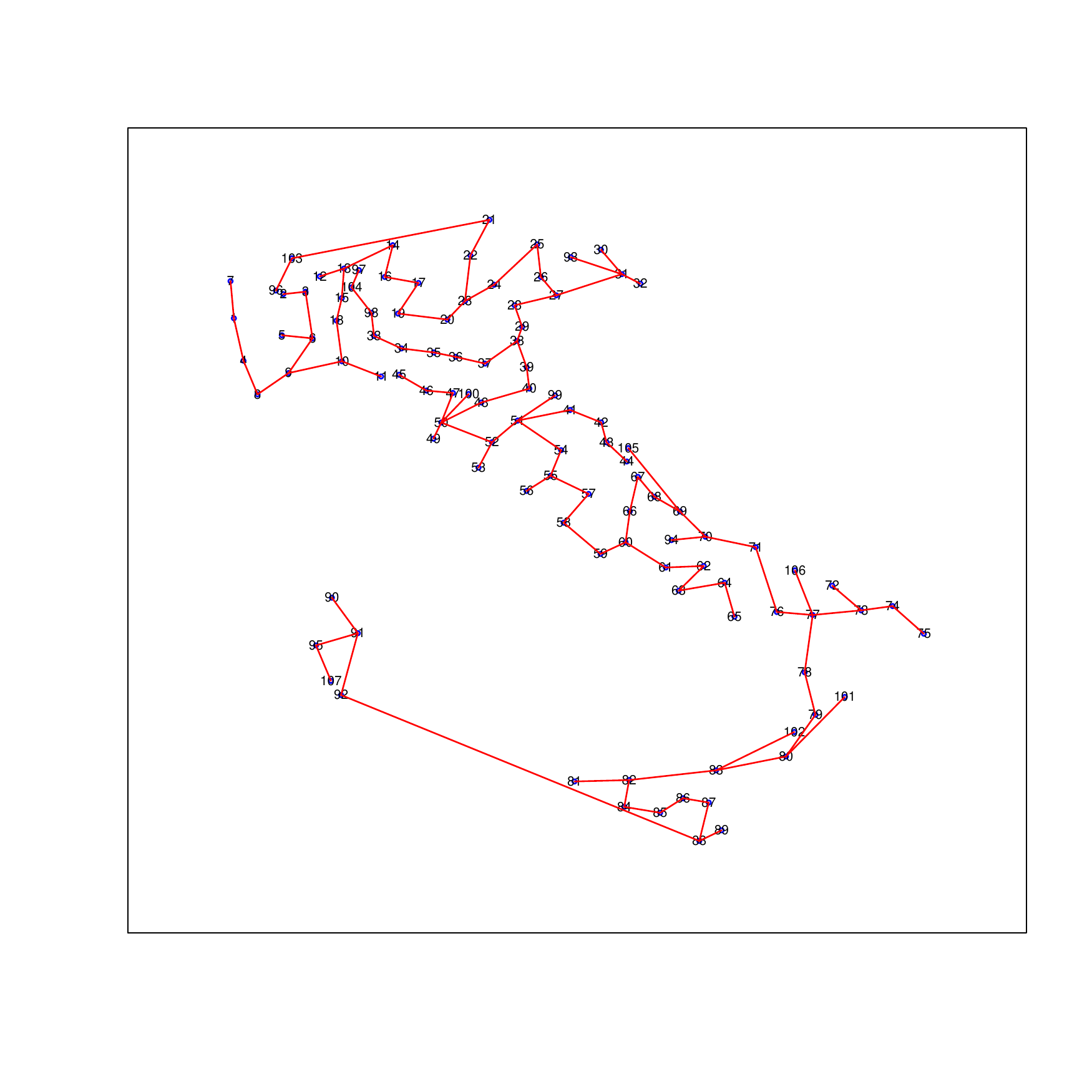}
\caption{Provincial minimum spanning neighbourhood graph}
\label{PROV_PGM4_vicinato}
\end{figure}

\begin{figure}[!htbp]
\centering
\includegraphics[width=.60\columnwidth]{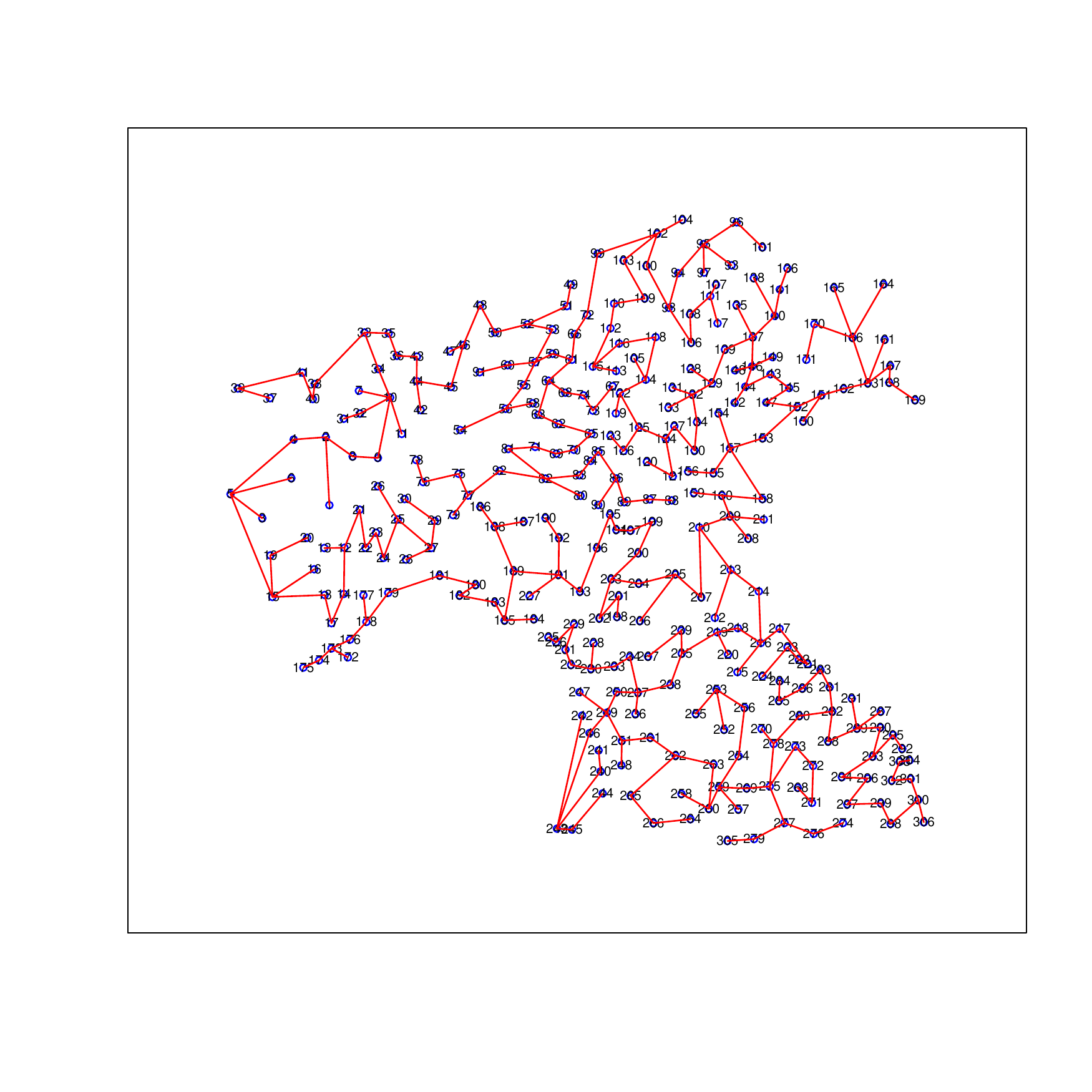}
\caption{LMAs minimum spanning neighbourhood graph}
\label{SSL_PGM4_vicinato}
\end{figure}


\newpage
\footnotesize
\bibliographystyle{elsarticle-harv}
\bibliography{z_paper}

\end{document}